\documentclass[onecolumn,tightenlines,nofootinbib,preprintnumbers,amsmath,amssymb]{revtex4}

\usepackage{slashed} 

\usepackage[usenames,dvipsnames]{color}
\usepackage{dcolumn}
\usepackage{graphicx}
\usepackage{epstopdf}
\usepackage{bm}
\usepackage{booktabs}
\usepackage{lscape}
\usepackage{tikz}
\usepackage[caption=false]{subfig} 

\usepackage{hyperref}
\usepackage{cleveref}
\renewcommand{\arraystretch}{0.9}
\usepackage{setspace}

\begin{document}
	\begin{spacing}{1.5}

\title{
Localized CP violation driven by cross-channel interference between $\overline{K}_0^*(1430)^0$ and $\rho(770)^0$ in $B^{\pm}\rightarrow K^{\pm}\pi^{+}\pi^{-}$ decays }

\author{Jin-Zhao Guo $^{1}$, Gang L\"{u}$^{1}$\footnote{Email: ganglv66@sina.com}}

\affiliation{\small $^{1}$School of Physics and Advanced Energy, Henan University of Technology, Zhengzhou 450001, China\\
}

\begin{abstract}
For the charmless three-body decay $B^{\pm} \rightarrow K^{\pm} \pi^{+} \pi^{-}$, a CP asymmetry sign reversal typically occurs in the low-$m_{\pi\pi}$ region. However, experimental data reveals a localized anomaly in the specific phase space region $0.4 \text{ GeV}^2 < m_{\pi\pi}^2 < 0.8 \text{ GeV}^2$ and $1.5 \text{ GeV}^2 < m_{K\pi}^2 < 2.5 \text{ GeV}^2$, where the regular pattern is broken by a rightward tilt. In this work, we investigate the dynamical origin of this phenomenon by employing a quasi-two-body factorization scheme that integrates short-distance perturbative QCD (PQCD) hard kernels with long-distance $S$-wave LASS and $P$-wave relativistic Breit--Wigner(RBW) line shapes. Our results demonstrate that the localized anomaly is driven by cross-channel interference between the scalar $\overline{K}_0^*(1430)^0$ and vector $\rho(770)^0$ resonances, which exhibit an orthogonal geometric topology on the Dalitz plot. Crucially, we show that the rapid variation of the relative strong phase within the intersection region induces coherent interference effects alternating between the charge-conjugated channels, significantly reshaping the local CP asymmetry profile. This cross-channel dynamics framework provides a self-consistent theoretical benchmark for overlapping resonance regions, offering valuable insights for upcoming high-luminosity experiments at Belle II and the HL-LHC.
\end{abstract}

\maketitle

\section{Introduction}
\label{sec:sample1}
				
Charmless three-body non-leptonic decays of heavy $B$ mesons serve as critical channels for testing the Standard Model (SM) of particle physics, searching for new physics signals beyond the SM, and precisely understanding the non-perturbative dynamical mechanisms of strong interactions~\cite{Cheng:2007shb,Klein:2017xti}. Due to the rich intermediate resonances and complex non-resonant backgrounds involved, these multi-body decay processes exhibit extremely rich hadronic interactions and local charge-conjugation and parity (CP) violation (CPV) behaviors in the two-dimensional total phase space Dalitz plot~\cite{Belle:2005rpz,LHCb:2019sus,BaBar:2008lpx,Belle:2004drb,LHCb:2013fio}.

In phase space, the quantum coherent superposition among different intermediate resonances can induce a drastic flipping of the strong interaction phase, thereby driving the non-smooth evolution of local CP violation across the phase space~\cite{Cheng:2020iwx}. For instance, in the classic $B^{\pm} \rightarrow K^{\pm} \pi^{+} \pi^{-}$ and $B^{\pm} \rightarrow \pi^{\pm} \pi^{+} \pi^{-}$ decay channels~\cite{LHCb:2022fpg,LHCb:2019jta}, the same-channel coherent interference between the vector resonance $\rho(770)^0$ and the scalar resonance $f_0(500)$ induces a striking sign reversal of the local CP asymmetry along the $\pi^+\pi^-$ invariant-mass axis near the $\rho(770)^0$ region. Within this narrow mass region, the local CP asymmetry sweeps across a wide range from $+0.7$ to $-0.7$. Such a sharp dynamical fluctuation not only helps us clearly capture the microscopic trajectory of the rapid strong phase evolution with the invariant mass, but also effectively breaks the mathematical degeneracy between the strong and weak phases, thereby achieving a highly precise experimental extraction of the Cabibbo--Kobayashi--Maskawa (CKM) weak phase angle $\gamma$~\cite{Bhattacharya:2013boa,Cabibbo:1963yz,Giri:2003ty}.

\begin{figure}[htbp]
	\centering

	\subfloat[Asymmetry distribution in bins of the Dalitz plot for ${B^{ \pm} \rightarrow K^{ \pm} \pi^{+} \pi^{-}}$\label{fig:dalitz_full}]{%
		\begin{tikzpicture}
			
			\node[anchor=south west, inner sep=0] (myimg) {\includegraphics[width=0.48\textwidth, keepaspectratio]{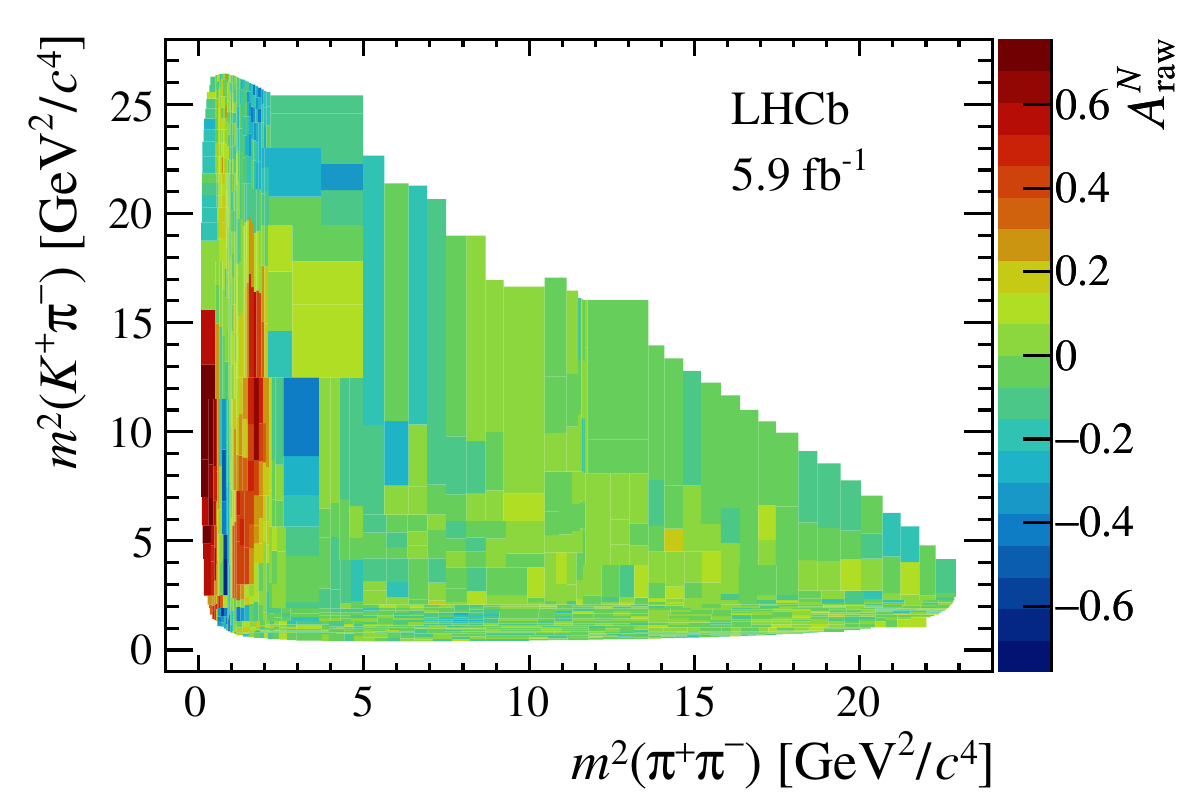}};

			\begin{scope}[x={(myimg.south east)}, y={(myimg.north west)}]

				\draw[red, thick] (0.175, 0.225) rectangle (0.195, 0.26);
				
			\end{scope}
		\end{tikzpicture}%
	}
	\hspace{0.4cm}%
	\subfloat[Zoomed-in view of the low-mass interference region\label{fig:dalitz_zoom}]{%
		\includegraphics[width=0.36\textwidth, trim=3.3cm 3cm 16.1cm 10cm, clip]{Fig3b.pdf}%
	}

	\caption{(a) The full phase space distribution with the low-mass region indicated by red solid box; (b) The corresponding zoomed-in view of the cross-channel interference region.}
	\label{fig:dalitz_combined}
\end{figure}

Figure~\ref{fig:dalitz_combined} displays the latest high-statistics Dalitz plot data of the $B^{\pm} \rightarrow K^{\pm} \pi^{+} \pi^{-}$ decay from the LHCb Collaboration~\cite{LHCb:2022fpg}, where the red solid box indicates a localized small phase space region defined by $0.4\,\mathrm{GeV}^2 < m_{\pi\pi}^{2} < 0.8\,\mathrm{GeV}^{2}$ and $1.5\,\mathrm{GeV}^2 < m_{K\pi}^{2} < 2.5\,\mathrm{GeV}^{2}$. Since the interference pattern between the $\pi\pi$ $S$-wave (including $f_0(500)$) and $P$-wave $\rho(770)^0$ components is governed by the $\pi\pi$ invariant-mass squared axis (with $s_{\pi\pi}$ as the independent variable) while the $K\pi$ axis (with $s_{K\pi}$ as the independent variable) only modulates the helicity angle factor, variation of $s_{K\pi}$ within this localized region does not induce any rapid fluctuations in the CP asymmetry. 
Consequently, the local CP dynamics should theoretically be dominated by the $s_{\pi\pi}$ dimension under single-channel frameworks. However, this region does not exhibit the typical CP sign-reversal pattern predicted by single-channel dynamical models. Instead, as clearly observed in Fig.~\ref{fig:dalitz_combined}(b), the CP asymmetry on the right-hand side exhibits rapid fluctuations significantly modulated by the variation of $s_{K\pi}$. We argue that this anomalous phenomenon is precisely attributed to the strong quantum coherent superposition between the scalar resonance $\overline{K}_0^*(1430)^0$ (hereafter denoted as $\overline{K}_0^{*0}$) located at the scalar pole $m_{K\pi}^{2} \approx 2.05\,\mathrm{GeV}^{2}$ and the vector resonance $\rho(770)^0$ (hereafter denoted as $\rho^0$). In the global dynamics of the three-body decay $B^{\pm} \rightarrow K^{\pm} \pi^{+} \pi^{-}$, the scalar resonance $\overline{K}_0^{*0}$ and the vector resonance $\rho^0$ belong to entirely distinct final-state subsystem combinations. Since these two resonance axes present a mutually orthogonal geometric topology on the two-dimensional Dalitz plot, the quantum coherent superposition induced in their intersection region essentially belongs to a typical cross-channel interference~\cite{Dedonder:2010fg}.

However, for a long period in the past, cross-channel interference received relatively little attention from both experimental and theoretical perspectives. From the standpoint of experimental history, charmless three-body $B$ meson decays are inherently suppressed by the multi-body decay chains, resulting in small total branching fractions, typically of $\mathcal{O}(10^{-5})$. Coupled with the limited luminosity of colliders in the early years, the accumulated event yields were far lower than what is available today~\cite{Belle:2005rpz, BaBar:2008lpx, HFLAV:2022pvk}. During the era of the first-generation $B$ factories (such as BaBar and Belle), the total signal yields for this specific channel were severely limited by the integrated luminosity, typically confined to a few thousand. Consequently, within the narrow cross-channel intersection region where the scalar and vector resonance bands overlap, only a handful of events fell into this extremely restricted two-dimensional area. This severe scarcity rendered the unique quantum coherent fringes of cross-channel interference highly susceptible to being entirely obscured by large statistical fluctuations. In recent years, as the LHCb experiment enters a high-statistics boom era, tens of thousands or even hundreds of thousands of clean three-body events make it possible to disentangle the fine hadronic structure of such an extremely narrow intersection region with high resolution, which also presents an urgent need for precise calculations and phenomenological predictions from the theoretical community~\cite{Wang:2020gqm,Zhou:2023dcl,Chai:2022prb}.

\section{Quasi-Two-Body Factorization and Resonance Line Shapes}\label{Sec 2}
				
In the Dalitz plot, the intersection of the $\overline{K}_0^{*0}$ and $\rho^{0}$ resonance bands along the $K\pi$ and $\pi\pi$ invariant mass axes leads to cross-channel interference. To describe this coherent superposition and provide a theoretical foundation for the experimental Isobar model, we adopt a quasi-two-body factorization scheme within the pole-approximated framework~\cite{Hua:2020usv,Chen:2002th,Zou:2020fax}. In this approach, the continuous multi-body scattering process is decomposed into two stages: short-distance weak production and long-distance final-state interaction.

Under this framework, the short-distance weak production processes reduce to the full quasi-two-body decay amplitudes, $\mathcal{A}_{S}$ and $\mathcal{A}_{V}$, which describe the localized prompt transitions $B^{-} \to \overline{K}_0^{*0}\pi^{-}$ and $B^{-} \to \rho^0 K^{-}$, respectively. To incorporate subsequent long-range final-state interactions, the $S$-wave amplitude $\mathcal{M}_{S}(s_{K\pi})$ for the three-body decay $B^{-} \rightarrow (\overline{K}_0^{*0} \rightarrow K^{-} \pi^{+}) \pi^{-}$ is expressed as the product of the full quasi-two-body amplitude $\mathcal{A}_{S}$ and the LASS-parameterized continuous spectrum of the $K\pi$ system:
\begin{align}
	\mathcal{M}_{S}(s_{K\pi}) = \mathcal{A}_{S} \times \frac{q_0}{m_0^2 \Gamma_0} g_{K_0^{*}} R(s_{K\pi}),
		\label{11}
\end{align}
where $s_{K\pi}$ is the invariant mass squared of the $K\pi$ system, $m_0$ and $\Gamma_0$ denote the pole mass and nominal total decay width of the $\overline{K}_0^{*0}$ resonance, respectively. Here, $q_0$ is the on-shell momentum evaluated at $s_{K\pi} = m_0^2$, with the expression of $q$ given in Eq.~\eqref{55}. The kinematic factor $\frac{q_0}{m_0^2 \Gamma_0}$ ensures the correct dimensionality and regulates the high-energy behavior of the non-resonant background. Taking into account the isospin Clebsch-Gordan coefficient for the decay $\overline{K}_0^{*0} \to K^- \pi^+$ (where the partial decay width satisfies $\Gamma(\overline{K}_0^{*0} \to K^- \pi^+) = \frac{2}{3}\Gamma_0$), the effective strong decay coupling constant, defined via the matrix element $g_{K_0^*} = \langle K^- \pi^+ | \overline{K}_0^{* 0} \rangle$, is explicitly given by~\cite{Cheng:2013dua}:
\begin{align}
	g_{K_{0}^{*}} = \sqrt{\frac{8 \pi m_{0}^{2} (\frac{2}{3}\Gamma_{0})}{q_{0}}}.
\end{align}
Following the standard convention in quasi-two-body factorization frameworks, the on-shell effective coupling $g_{K_0^*}$ is defined to be real and positive. In contrast to empirical parameterizations where arbitrary strong phases are attached to effective couplings (e.g., Ref.~\cite{Cheng:2020iwx}), here all physical strong phase variations are dynamically governed by the short-distance PQCD kernels and the long-distance unitary line shapes.

The phenomenological scattering amplitude $R(s_{K\pi})$ characterizes the continuous spectrum of the $S$-wave $K\pi$ state and consists of an elastic scattering background and a resonant component~\cite{Belle:2019rup,Aston:1987ir,BaBar:2005qms}:
\begin{align}
	R(s_{K\pi}) = \frac{\sqrt{s_{K\pi}}}{q \cot\delta_B - i q} + e^{2i\delta_B} \frac{\frac{m_0^2 \Gamma_0}{q_0}}{m_0^2 - s_{K\pi} - i m_0 \Gamma_0 \left(\frac{q}{\sqrt{s_{K\pi}}}\right) \frac{m_0}{q_0}}.
\end{align}
The first term represents the slowly varying non-resonant background, governed by the scattering length $a$ and the effective range $r$. The phase shift $\delta_B$ satisfies:
\begin{align}
	\cot\delta_B = \frac{1}{a q} + \frac{1}{2} r q.
\end{align}
Following experimental results, these empirical parameters are fixed at $a = 2.07 \pm 0.10 \text{ GeV}^{-1}$ and $r = 3.32 \pm 0.34 \text{ GeV}^{-1}$~\cite{BaBar:2005qms}. The second term is the RBW line shape for the $\overline{K}_0^{*0}$ resonance. The variable $q$ is the momentum magnitude of the daughter mesons in the $K\pi$ rest frame~\cite{Wang:2020saq}:
\begin{align}
	q = \frac{1}{2}\sqrt{\frac{[s_{K\pi}-(m_K+m_\pi)^2][s_{K\pi}-(m_K-m_\pi)^2]}{s_{K\pi}}},
	\label{55}
\end{align}
where $m_K$ and $m_\pi$ denote the masses of the $K$ and $\pi$ mesons, respectively. The phase shift factor $e^{2i\delta_B}$ ensures partial-wave unitarity across the phase space~\cite{Wang:2020saq}, while the full short-distance hadronic inputs from the transitions into scalar and pseudoscalar states remain consistently embedded within the quasi-two-body amplitude $\mathcal{A}_{S}$.

Because the hard scattering matrix elements vary slowly with momentum within the localized intersection region, we approximate them by freezing the kinematic variables at the resonance pole ($s_{K\pi} = m_0^2$). This approach isolates $\mathcal{A}_{S}$, which represents the full quasi-two-body amplitude for the $B^{-} \rightarrow \overline{K}_0^{*0} \pi^{-}$ transition, from the running phase space, treating it as a constant complex coefficient across the entire Dalitz plot. The validity of this pole-freezing approximation is well-supported by systematic calculations in Ref.~\cite{Wang:2020saq}, which demonstrate that neglecting the $s$-dependence in the hard kernels introduces a discrepancy of only about $10\%$ in the factorization relation for $S$-wave $B \to K_0^*(1430)\pi$ decays.
Such a factorization decouples the perturbative quark-gluon kernels from the long-distance wave function evolution, thereby preserving the physical line shape within the form factor. Similarly, for the $P$-wave vector sector governed by the $\rho^0$ resonance, we apply the same quasi-two-body factorization scheme by freezing the short-distance weak kernel at the vector meson pole ($s_{\pi\pi} = m_\rho^2$). The three-body decay amplitude mediated by the $\rho^0$ resonance channel is formulated as the product of the constant two-body amplitude $\mathcal{A}_{V}$ and the time-like form factor of the $\pi\pi$ system~\cite{Zhang:2013oqa,Zhang:2013iga}:
\begin{align}
	\mathcal{M}_{V}(s_{\pi\pi}, \theta_H) = \mathcal{A}_{V}\times \frac{\mathcal{F}_{\pi\pi}(s_{\pi\pi}, \theta_H)}{p_{B} \cdot \epsilon_L^{*}},
\end{align}
where $s_{\pi\pi}$ is the invariant mass squared of the $\pi^+\pi^-$ pair, and $\theta_H$ is the helicity angle between the daughter pion and the bachelor kaon. Here, $\mathcal{A}_{V}$ represents the full longitudinal two-body weak decay amplitude, and $\epsilon_L^*$ denotes the longitudinal polarization vector of the intermediate $\rho$ meson.  The $P$-wave time-like form factor $\mathcal{F}_{\pi\pi}(s_{\pi\pi}, \theta_H)$ incorporates the strong decay vertex, the resonant line shape, and the angular momentum barrier effects, parameterized via a generalized RBW form~\cite{Wang:2015ula}:
\begin{align}
	\mathcal{F}_{\pi\pi}(s_{\pi\pi}, \theta_H) = \frac{-2 g_{\rho} |\mathbf{p}_2||\mathbf{p}_3|\cos\theta_H}{s_{\pi\pi} - m_\rho^2 + i m_\rho \Gamma_\rho},
\end{align}
where $m_\rho$ and $\Gamma_\rho$ denote the pole mass and the nominal decay width of the $\rho^0$ resonance, respectively, and $g_{\rho}$ is the effective strong coupling constant for the $\rho^0 \to \pi^+\pi^-$ decay. The variable $\theta_H$ denotes the helicity angle of the $\pi^+\pi^-$ subsystem, defined specifically as the angle between the momentum of the daughter meson $\pi^+$ and the momentum of the bachelor meson $K^-$ evaluated in the rest frame of the intermediate vector resonance $\rho^0$. The kinematic factor $-2\vert{}\mathbf{p}_2\vert{}\vert{}\mathbf{p}_3\vert{}\cos\theta_H$ in the numerator originates from the Lorentz tensor contraction of the vector polarization sum, accounting for the centrifugal barrier of the $P$-wave state~\cite{Wang:2015ula, Zhang:2013oqa}. Here, $\vert{}\mathbf{p}_2\vert{}$ and $\vert{}\mathbf{p}_3\vert{}$ represent the momentum magnitudes of the daughter pion and the bachelor kaon, respectively, both evaluated in the $\pi^+\pi^-$ rest frame~\cite{Bediaga:2006jk}.

Here, we employ a constant nominal width $\Gamma_\rho$ in the $\rho^0$ propagator. While a running width is critical for unitarizing the $S$-wave $K\pi$ spectrum near the threshold, a constant approximation is highly self-consistent for the $P$-wave $\rho^0$ channel, since the studied region ($0.4\text{ GeV}^2 < s_{\pi\pi} < 0.8\text{ GeV}^2$) is tightly localized around the $\rho$ pole and sits far above the kinematic threshold ($4m_\pi^2 \approx 0.08\text{ GeV}^2$). This choice keeps the model analytically clean by avoiding unconstrained barrier radius parameters, allowing us to isolate the net cross-channel interference.

Crucially, inside such an exceedingly restricted phase space region ($0.4\text{ GeV}^2 < s_{\pi\pi} < 0.8\text{ GeV}^2$ and $1.5\text{ GeV}^2 < s_{K\pi} < 2.5\text{ GeV}^2$), the dynamic behavior of the total amplitude is heavily dominated by the rapid running of the strong phases within the resonance propagators, whereas the phase-space variation of the angular distribution factor $\cos\theta_H$ is kinematically suppressed. Across the physically allowed domain of this intersection region, $\cos\theta_H(s_{\pi\pi}, s_{K\pi})$ varies smoothly and monotonically within a narrow range from $-1.0$ (at the lower Dalitz boundary) to $-0.85$ (near the upper boundary). Relative to its evaluation at the resonance poles ($s_{\pi\pi}=m_\rho^2, s_{K\pi}=m_0^2$), where $\cos\theta_H \approx -0.93$, this mild variation ($\le 8\%$) without zero-crossings or sign flips exerts no decisive topological influence on the overall interference pattern. In alignment with the localized, model-independent binning frameworks typically executed in experimental phase-space analyses~\cite{LHCb:2022fpg}, it is therefore well-justified and self-consistent to fix $\cos\theta_H$ to its effective pole value of $-0.93$, treating it as an effective constant factor across this core region.

\section{Total decay amplitude in the PQCD framework}\label{Sec 3}

Conforming to the pole-approximated reduced amplitude scheme established in Sec.~\ref{Sec 2}, the short-distance micro-dynamics of the charmless non-leptonic decays are isolated by freezing the kinematic running variables exactly at the intermediate resonance poles. Consequently, the core perturbative quark-gluon matrix elements across the overlapping region self-consistently simplify to the evaluation of the effective two-body reduced decay amplitudes, $\mathcal{A}_{S}$ and $\mathcal{A}_{V}$. In this section, we formulate these hard scattering kernels within the framework of the perturbative QCD (PQCD) approach based on the $k_T$ factorization theorem.

The three-body decay process is accompanied by multifaceted dynamical mechanisms. The perturbative QCD (PQCD) method, originating from the pioneering works based on the $k_T$ factorization theorem~\cite{Keum:2000wi,Lu:2000em,Chen:2002th}, is known for its efficacy in handling perturbation corrections. It has been successfully applied to a wide range of exclusive two-body non-leptonic decay processes including comprehensive next-to-leading order (NLO) corrections~\cite{Cheng:2014pqa, Xiao:2022ebt}, and holds promise for quasi-two-body decay processes as well. In the framework of PQCD, within the rest frame of a heavy $B$ meson, the decay involves the production of two highly energetic light mesons. The dominance of hard interactions arises from the suppression of soft gluon exchanges between the fast-recoiling final states~\cite{Li:1994iu}, wherein a hard gluon imparts momentum to the spectator quark, mediated by six-quark operators~\cite{Keum:2000ph}. Crucially, within the studied phase-space region ($s_{K\pi} < 2.5\text{ GeV}^2$, corresponding to $m_{K\pi} \lesssim 1.58\text{ GeV}$), the recoiling bachelor pion carries a large energy of $E_\pi \approx 2.4\text{ GeV} \gg \Lambda_{\text{QCD}}$. This ample energy release ensures that the characteristic hard scale $\mu \sim \mathcal{O}(\sqrt{\bar{\Lambda} m_B}) \sim 1.5\text{--}2.0\text{ GeV}$ remains within the perturbative regime, justifying the factorization expansion.

Within the quasi-two-body framework, these multi-body transitions are evaluated by freezing the hard scattering kernels at the intermediate resonance poles, decoupling short-distance quark interactions from long-distance meson line shapes. More explicitly, in the generalized multi-body PQCD scheme, the non-perturbative dynamics of the $S$-wave $K\pi$ pair is encapsulated within the two-meson distribution amplitude (2MDA), normalized by the time-like scalar form factor. Following Ref.~\cite{Wang:2020saq}, embedding the experimental LASS continuous spectrum corresponds to replacing this form factor via $\hat{R}(s) = \frac{q_0}{m_0^2 \Gamma_0} g_{K_0^*} \bar{f}_{K_0^*} R(s)$. Under the pole-freezing approximation, the scalar decay constant $\bar{f}_{K_0^*}$ is absorbed into the standard single-meson distribution amplitude (1MDA) to evaluate the short-distance amplitude $\mathcal{A}_S$, while the energy-dependent factor $\frac{q_0}{m_0^2 \Gamma_0} g_{K_0^*} R(s_{K\pi})$ is factored out as formulated in Eq.~\eqref{11}. Systematic evaluations in Ref.~\cite{Wang:2020saq} demonstrate that this quasi-two-body reduction introduces a discrepancy of around 10\% compared to the full 2MDA convolution, providing both computational tractability and theoretical consistency.

To describe the intersection region where the $S$-wave $\overline{K}_0^{*0}$ and $P$-wave $\rho^0$ states overlap, the total scattering amplitude must be formulated coherently. Within the quasi-two-body factorization framework, this total three-body decay amplitude $\mathcal{M}(s_{\pi\pi}, s_{K\pi})$ is constructed through the coherent superposition of the $S$-wave scalar sector ($B^{-} \rightarrow (\overline{K}_0^{*0} \rightarrow K^{-} \pi^{+}) \pi^{-}$) and the $P$-wave vector sector ($B^{-} \rightarrow (\rho^{0} \rightarrow \pi^{-} \pi^{+}) K^{-}$):
\begin{align}
	\mathcal{M}(s_{\pi\pi}, s_{K\pi}) = \mathcal{M}_{S}(s_{K\pi}) + \mathcal{M}_{V}(s_{\pi\pi}, \theta_H).
\end{align}
This approach self-consistently embeds the short-distance weak production kernels calculated from the six-quark operators into the long-distance continuous resonant line shapes. Incorporating the localized prompt reduced amplitudes $\mathcal{A}_S$ and $\mathcal{A}_V$ with their respective time-like form factors and CKM structures, the fully expanded total decay amplitude across the phase space is explicitly formulated as follows:

\begin{equation}
	\begin{split}
		\mathcal{M}(s_{\pi\pi}, s_{K\pi}) = &\frac{q_{0} g_{K_{0}^{*}} R(s_{K\pi})}{m_{0}^{2}\Gamma_{0}} \times \bigg[ \frac{G_F}{\sqrt{2}} V_{ub}V_{us}^* \Big\{ a_1 \mathcal{A}_{ef}^{LL}(\overline{K}_0^{*0}, \pi) + C_1 \mathcal{A}_{gh}^{LL}(\overline{K}_0^{*0}, \pi) \Big\} \\
		& - \frac{G_F}{\sqrt{2}} V_{tb}V_{ts}^* \Big\{ (a_4 - \frac{1}{2}a_{10})\mathcal{A}_{ab}^{LL}(\overline{K}_0^{*0}, \pi) + (a_6 - \frac{1}{2}a_{8})\mathcal{A}_{ab}^{SP}(\overline{K}_0^{*0}, \pi) \\
		& + (a_4 + a_{10})\mathcal{A}_{ef}^{LL}(\overline{K}_0^{*0}, \pi) + (a_6 + a_{8})\mathcal{A}_{ef}^{SP}(\overline{K}_0^{*0}, \pi) \\
		& + (C_3 - \frac{1}{2}C_9)\mathcal{A}_{cd}^{LL}(\overline{K}_0^{*0}, \pi) + (C_5 - \frac{1}{2}C_7)\mathcal{A}_{cd}^{SP}(\overline{K}_0^{*0}, \pi) \\
		& + (C_3 - \frac{1}{2}C_9)\mathcal{A}_{gh}^{LL}(\overline{K}_0^{*0}, \pi) + (C_5 - \frac{1}{2}C_7)\mathcal{A}_{gh}^{SP}(\overline{K}_0^{*0}, \pi) \Big\} \bigg] \\
		& +\frac{\mathcal{F}_{\pi\pi}(s_{\pi\pi}, \theta_H)}{p_{B} \cdot \epsilon_L^{*}} \times \bigg[ \frac{G_{\mathrm{F}}}{2} V_{u b} V_{u s}^{*}\Big\{a_{1}\big[\mathcal{A}_{a b}^{L L}(\overline{K}, \rho)+\mathcal{A}_{e f}^{L L}(\overline{K}, \rho)\big] \\
		& +a_{2} \mathcal{A}_{a b}^{L L}(\rho, \overline{K})+C_{2}\big[\mathcal{A}_{c d}^{L L}(\overline{K}, \rho)+\mathcal{A}_{g h}^{L L}(\overline{K}, \rho)\big]+C_{1} \mathcal{A}_{c d}^{L L}(\rho, \overline{K})\Big\} \\
		& -\frac{G_{\mathrm{F}}}{2} V_{t b} V_{t s}^{*}\Big\{\left(a_{4}+a_{10}\right)\big[\mathcal{A}_{a b}^{L L}(\overline{K}, \rho)+\mathcal{A}_{e f}^{L L}(\overline{K}, \rho)\big] \\
		& +\left(a_{6}+a_{8}\right)\big[\mathcal{A}_{a b}^{S P}(\overline{K}, \rho)+\mathcal{A}_{e f}^{S P}(\overline{K}, \rho)\big]+\frac{3}{2}\left(a_{7}+a_{9}\right) \mathcal{A}_{a b}^{L L}(\rho, \overline{K}) \\
		& +\left(C_{3}+C_{9}\right)\big[\mathcal{A}_{c d}^{L L}(\overline{K}, \rho)+\mathcal{A}_{g h}^{L L}(\overline{K}, \rho)\big] \\
		& +\left(C_{5}+C_{7}\right)\big[\mathcal{A}_{c d}^{S P}(\overline{K}, \rho)+\mathcal{A}_{g h}^{S P}(\overline{K}, \rho)\big] \\
		& +\frac{3}{2} C_{8} \mathcal{A}_{c d}^{L R}(\rho, \overline{K})+\frac{3}{2} C_{10} \mathcal{A}_{c d}^{L L}(\rho, \overline{K})\Big\}\bigg],
	\end{split}
\end{equation}
where $G_{\mathrm{F}}$ denotes the Fermi coupling constant, and the superscripts $LL$, $LR$, and $SP$ represent the three types of chiral current structures. The subscripts $ab, cd, ef$, and $gh$ characterize the specific topologies of the hard scattering Feynman diagrams: $\mathcal{A}_{ab}$ and $\mathcal{A}_{cd}$ denote the factorizable and nonfactorizable emission diagrams, respectively, while $\mathcal{A}_{ef}$ and $\mathcal{A}_{gh}$ correspond to the factorizable and nonfactorizable annihilation diagrams, respectively. The explicit expressions for these amplitudes are detailed in Ref.~\cite{Yang:2022ebu}.
The coefficients $C_i$ ($i = 1, 2, \dots, 10$) represent the standard Wilson coefficients~\cite{Ali:1998eb,Buchalla:1995vs}. The CKM matrix, whose elements are determined through experimental 
observations, can be expressed in terms of the Wolfenstein parameters $A, \rho, \lambda$, and $\eta$ : $V_{tb} V_{ts}^{*} = -A \lambda^{2}, V_{u b} V_{u s}^{*}=A \lambda^{4}(\rho-i \eta)$. The effective coefficients $a_i$ are composed of combinations of the Wilson coefficients and the color factor $N_c$, characterizing the vertex properties of specific topological diagrams. Specifically, the effective coefficients $a_i$ obey the following relations:
\begin{equation}
	\begin{split}
a_{i} = \left\{ \begin{array}{ll}
	C_{i} + \frac{1}{N_{c}} C_{i+1}, & \text{for odd } i, \\
	C_{i} + \frac{1}{N_{c}} C_{i-1}, & \text{for even } i.
\end{array} \right.
	\end{split}
\end{equation}

In the differential decay rate $|\mathcal{M}(s_{\pi\pi},s_{K\pi})|^2$, the cross-channel interference term $2\text{Re}[\mathcal{M}_S\mathcal{M}_V^*]$ arises, where $\text{Re}$ denotes the real part of the complex expression enclosed in the brackets. This interference term is governed by the relative strong phase difference:
\begin{equation}
	\Delta\delta(s_{\pi\pi},s_{K\pi}) = [\delta_S^{\text{hard}} + \delta_S^{\text{line}}(s_{K\pi})] - [\delta_V^{\text{hard}} + \delta_V^{\text{line}}(s_{\pi\pi})],
\end{equation}
along with the CKM weak phase structures. Here, $\delta_{S,V}^{\text{hard}}$ denote the short-distance dynamical strong phases perturbatively generated by the PQCD hard kernels (arising from the imaginary parts of loop, vertex, and annihilation diagrams), while $\delta_S^{\text{line}}(s_{K\pi}) = \text{Arg}[R(s_{K\pi})]$ and $\delta_V^{\text{line}}(s_{\pi\pi}) = \text{Arg}[\mathcal{F}_{\pi\pi}(s_{\pi\pi})]$  represent the long-distance running phases determined by the unitary $S$-wave LASS and $P$-wave RBW line shapes, respectively. Thus, the phase variations across the phase space are self-consistently determined within the QCD factorization framework without introducing additional empirical relative phases.

To investigate the localized direct CP violation induced by this interference, we integrate over a specific phase space domain $\Omega$. The localized integrated direct CP asymmetry $A_{CP}^\Omega$ is defined as:
\begin{align}
	A_{\mathrm{CP}}^{\Omega} = \frac{\iint_{\Omega} \mathrm{d}s_{\pi\pi} \mathrm{d}s_{K\pi} \left( |\mathcal{M}(s_{\pi\pi}, s_{K\pi})|^2 - |\overline{\mathcal{M}}(s_{\pi\pi}, s_{K\pi})|^2 \right)}{\iint_{\Omega} \mathrm{d}s_{\pi\pi} \mathrm{d}s_{K\pi} \left( |\mathcal{M}(s_{\pi\pi}, s_{K\pi})|^2 + |\overline{\mathcal{M}}(s_{\pi\pi}, s_{K\pi})|^2 \right)},
\end{align}
where $\overline{\mathcal{M}}(s_{\pi\pi}, s_{K\pi})$ is the CP-conjugated total amplitude. The integration domain $\Omega$ is bounded by the physical Dalitz plot kinematics. For a localized region focused on the intersection core (e.g., $0.4 \text{ GeV}^2 < s_{\pi\pi} < 0.8 \text{ GeV}^2$ and $1.5 \text{ GeV}^2 < s_{K\pi} < 2.5 \text{ GeV}^2$), the integration is truncated at the physical threshold limits. Within this narrow region, the angular term is explicitly fixed to its effective pole value of $\cos\theta_H \approx -0.93$ as justified above. Consequently, the short-distance inputs encoded in $\mathcal{A}_{S,V}$ and the long-distance strong phases from the line shapes completely govern the spatial fluctuations of the CP asymmetry, providing a clean theoretical benchmark for direct comparison with experimental Dalitz plot distributions.

\section{Input Parameters and Analysis of Data Results}\label{sec:results}
\begin{table}[htbp]
	\centering
	\renewcommand{\arraystretch}{1.4} 
	\caption{The input parameters, resonance constants, and basic kinematic masses utilized in the localized CP asymmetry analysis~\cite{Shen:2006ms, Yang:2022ebu,Wang:2020saq,ParticleDataGroup:2024cfk}.}
	\label{tab:input_parameters}
	\vspace{0.2cm}
	\begin{tabular}{llll}
		\hline
		\hline
				$m_B = 5.28 \text{ GeV}$ & $m_K = 0.494 \text{ GeV}$ & $m_\pi = 0.140 \text{ GeV}$ & $m_0 = 1.425 \text{ GeV}$ \\
		$m_\rho = 0.775 \text{ GeV}$ & $\Gamma_\rho = 0.149 \text{ GeV}$ & $g_\rho = 5.95$ &  $\Gamma_0 = 0.270 \text{ GeV}$ \\	$\omega_B = 0.40 \pm 0.04 \text{ GeV}$ & ~$m_0^\pi = 1.4 \pm 0.1 \text{ GeV}$ &~ $m_0^K = 1.6 \pm 0.1 \text{ GeV}$
 & \\
		$\lambda = 0.2250$ & $A = 0.826$ & $\bar{\rho} = 0.159$ & $\bar{\eta} = 0.352$ \\
		\hline
		\hline
	\end{tabular}
\end{table}

To implement the numerical analysis of the localized CP violation, the relevant inputs—including the fundamental kinematic masses, CKM Wolfenstein parameters, resonance constants, and hadronic distribution amplitude parameters—are consistently compiled in Table~\ref{tab:input_parameters}.
Before exploring the localized cross-channel interference dynamics in $B^\pm \to K^\pm \pi^+ \pi^-$, it is instructive to first examine the reliability of our PQCD calculations for the individual intermediate quasi-two-body decay channels. Using the input parameters specified in Table~\ref{tab:input_parameters}, we evaluate the branching ratios ($\mathcal{B}$) and direct $CP$ asymmetries ($A_{CP}$) for $B^{-} \to \overline{K}_{0}^{*}(1430)^{0}\pi^{-}$ and $B^{-} \to \rho(770)^{0}K^{-}$. The calculated values alongside the latest experimental measurements from the Particle Data Group (PDG)~\cite{ParticleDataGroup:2024cfk} are listed in Table~\ref{tab:individual_modes}.

\begin{table}[t]
	\centering
	\caption{Comparison of the calculated branching ratios ($\mathcal{B}$) and direct $CP$ asymmetries ($A_{CP}$) for individual intermediate quasi-two-body decay channels with experimental data from the Particle Data Group (PDG)~\cite{ParticleDataGroup:2024cfk}. The theoretical uncertainties of our results are classified into two categories: the first error comes from varying $\omega_B = 0.40 \pm 0.04$ GeV, and the second error originates from the chiral scaling mass of the final-state bachelor meson, with $m_0^\pi = 1.4 \pm 0.1$ GeV for the $\pi^-$ final state and $m_0^K = 1.6 \pm 0.1$ GeV for the $K^-$ final state~\cite{Shen:2006ms, Yang:2022ebu, Wang:2020saq}.}
	\label{tab:individual_modes}
	\renewcommand{\arraystretch}{1.25}
	\begin{ruledtabular}
		\begin{tabular}{lcccc}
			& \multicolumn{2}{c}{Branching Ratio ($\mathcal{B}$)} & \multicolumn{2}{c}{Direct $CP$ Asymmetry ($A_{CP}$)} \\
			\cline{2-3} \cline{4-5}
			\rule{0pt}{2.6ex}Decay Channel & This Work & PDG~\cite{ParticleDataGroup:2024cfk} & This Work & PDG~\cite{ParticleDataGroup:2024cfk} \\[0.5ex]
			\hline
			\rule{0pt}{2.6ex}$B^{-} \to \overline{K}_{0}^{*}(1430)^{0}\pi^{-}$ & $(4.27_{-0.62}^{+0.89}{}_{-0.19}^{+0.22}) \times 10^{-5}$ & $(3.9 \pm 0.4) \times 10^{-5}$ & $(-1.53_{-0.09}^{+0.11}{}_{-0.06}^{+0.04})\%$ & $(+4.4 \pm 2.4)\%$ \\
			$B^{-} \to \rho(770)^{0}K^{-}$                    & $(1.22_{-0.27}^{+0.36}{}_{-0.07}^{+0.09}) \times 10^{-6}$ & $(3.7 \pm 0.5) \times 10^{-6}$ & $(+65.63_{-3.37}^{+2.95}{}_{-2.12}^{+1.31})\%$ & $(+37 \pm 10)\%$ \\
		\end{tabular}
	\end{ruledtabular}
\end{table}

As shown in Table~\ref{tab:individual_modes}, the predicted branching fraction for $B^{-} \to \overline{K}_{0}^{*}(1430)^{0}\pi^{-}$ agrees very well with the experimental measurement within uncertainties, while its direct $CP$ asymmetry is confirmed to be suppressed near zero, in line with the pure penguin-dominated nature of this channel~\cite{Shen:2006ms}. 
For the $B^{-} \to \rho(770)^{0}K^{-}$ channel, our leading-order (LO) results reproduce the typical behavior documented in previous PQCD literature~\cite{Yang:2022ebu}. Specifically, while the predicted branching ratio is of the order of $\mathcal{O}(10^{-6})$ but lower than the PDG value by a factor of about three—a known tension in LO calculations that is expected to be alleviated by higher-order QCD radiative corrections~\cite{Yang:2022ebu}—our framework correctly captures the prominent positive sign and sizable magnitude of the direct $CP$ asymmetry observed experimentally. 
Overall, these benchmark evaluations demonstrate that our hadronic inputs and PQCD framework capture the essential short-distance dynamics consistently with existing literature, providing a reasonable leading-order foundation for analyzing the relative cross-channel interference phenomena on the Dalitz plot.

It is worth noting that the uncertainties from the electroweak CKM flavor-mixing parameters are not listed in Tables~\ref{tab:individual_modes} and \ref{tab:integrated_acp}. According to standard PQCD precision analyses, the variations induced by the modern PDG-constrained CKM parameters~\cite{ParticleDataGroup:2024cfk} are typically well within $1\%$. Such minor parametric uncertainties are completely dominated and obscured by the much larger non-perturbative hadronic uncertainties ($\omega_B$ and $m_0^{\pi,K}$) at the $10\%$ level. Therefore, the CKM uncertainties are safely neglected throughout our numerical evaluations.

The primary objective of this work is to elucidate the physical mechanisms underlying the localized CP-violating anomalies in the low-$s_{\pi\pi}$ and low-$s_{K\pi}$ regions. Within the scope of the present study, the $f_0(500)$ component is deliberately omitted from the total decay amplitude, a choice motivated by two explicit and rigorous physical arguments: 

Dynamically, while the same-channel S-P wave interference ($\rho^0 - f_0(500)$) primarily establishes the traditional sign-reversal baseline of the local CP asymmetry along the $s_{\pi\pi}$ axis---transitioning from positive at lower $s_{\pi\pi}$ to negative at higher $s_{\pi\pi}$---the pronounced topological distortion, specifically the rightward tilt observed within the intersection region ($0.4 \text{ GeV}^2 < s_{\pi\pi} < 0.8\text{ GeV}^2$ and $1.5 \text{ GeV}^2 < s_{K\pi} < 2.5\text{ GeV}^2$), is predominantly driven by the cross-channel $S$-$P$ wave coherent superposition between the vector resonance $\rho^0$ (in the $\pi\pi$ channel) and the scalar resonance $\overline{K}_0^{*0}$ (in the $K\pi$ channel). Consequently, omitting the $f_0(500)$ component serves as a necessary strategy to isolate and map the net dynamical effects of this novel cross-channel interference with maximum physical clarity.

Furthermore, from the perspective of theoretical model construction, excluding the $f_0(500)$ component avoids introducing severe model-dependent parameterization uncertainties. Due to its exceptionally broad width and strong entanglement with the non-resonant scattering background, the $f_0(500)$ cannot be parameterized by a simple isolated Breit-Wigner shape, and no compiled branching fraction is assigned in the Particle Data Group (PDG) summary tables~\cite{ParticleDataGroup:2024cfk}. Forcing a speculative parameterization of this diffuse background would introduce unconstrained hadronic uncertainties, thereby obscuring the extracted cross-channel signals. This deliberate isolation allows us to establish a clean and transparent theoretical benchmark exclusively for the $\overline{K}_0^{*0} \times \rho^0$ cross-channel interference network. In phenomenological studies aimed at identifying novel dynamic mechanisms, such a targeted isolation of the dominant interference term represents a well-justified and widely accepted approach.

Similarly, regarding the tensor resonance $K_2^*(1430)$, although observed in experimental multi-body analyses~\cite{ParticleDataGroup:2024cfk}, its contribution is omitted from our core framework based on two primary physical considerations. 
First, the localized $CP$ asymmetry sign reversal and anomalous tilt observed by LHCb span an extended phase-space window ($1.5\text{ GeV}^2 < s_{K\pi} < 2.5\text{ GeV}^2$, with $\Delta s_{K\pi} \approx 1.0\text{ GeV}^2$). This macroscopic topological profile is fundamentally governed by the continuous, long-range evolution of the relative strong phase $\Delta\delta(s_{\pi\pi}, s_{K\pi})$ provided by the broad $S$-wave LASS parameterization ($\Gamma_0 \approx 270\text{ MeV}$ alongside the extensive elastic phase shift $\delta_B(s_{K\pi})$). In contrast, the tensor state $K_2^*(1430)$ has a considerably narrower width ($\Gamma \approx 100\text{ MeV}$); its Breit-Wigner propagator attenuates rapidly away from the pole $s_{K\pi} \approx 2.04\text{ GeV}^2$, and its $D$-wave centrifugal barrier factor further suppresses the off-shell amplitude toward the kinematic boundaries. Consequently, while $K_2^*(1430)$ can induce localized fine modulations in the immediate vicinity of its pole, it cannot sustain the long-range strong phase variation required to drive the extended $1.0\text{ GeV}^2$ topological tilt. 
Second, isolating the $S$-$P$ wave cross-channel interference avoids introducing additional unconstrained hadronic parameters associated with tensor-meson distribution amplitudes and polarization configurations, thereby establishing a clean and transparent theoretical benchmark for the dominant cross-channel mechanism. 

\begin{figure}[htbp]
	\centering
	\includegraphics[height=9cm, width=9cm]{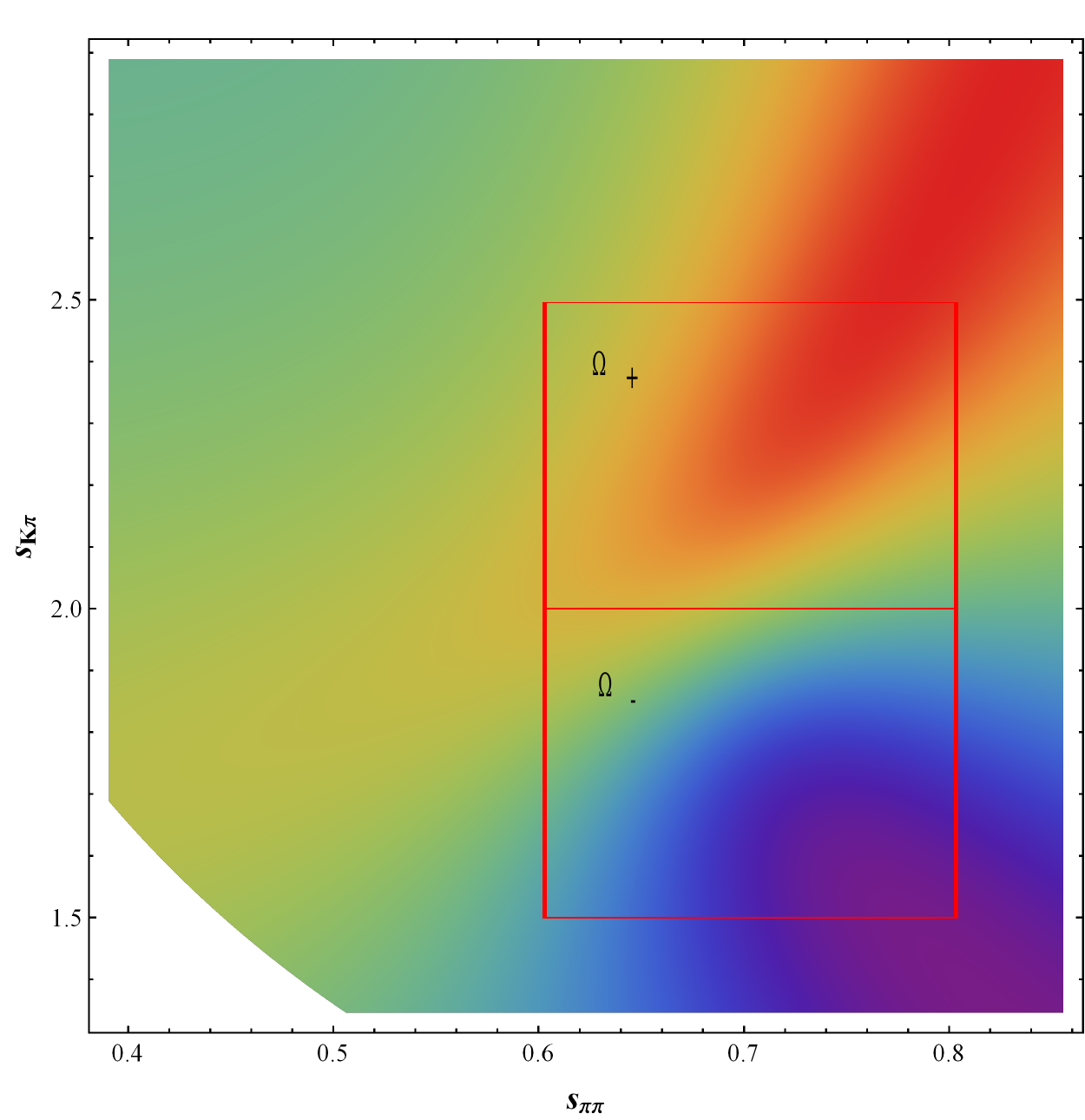}
	\hspace{0.3cm}
	\includegraphics[height=9cm, width=0.7cm]{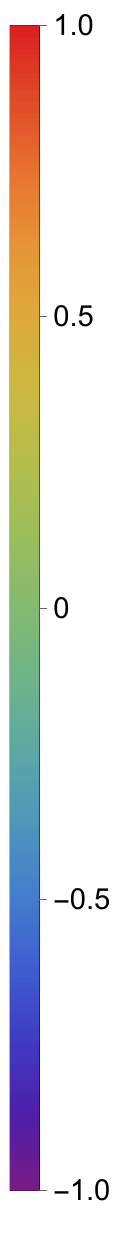}
	\caption{Localized $A_{\mathrm{CP}}$ distribution as a function of $s_{\pi\pi}$ and $s_{K\pi}$ in $B^{\pm} \rightarrow K^{\pm}\pi^{+}\pi^{-}$ decays, evaluated using the central values of input parameters. The plotted phase space focuses on the intersection region ($0.4\text{ GeV}^2 < s_{\pi\pi} < 0.8\text{ GeV}^2$, $1.5\text{ GeV}^2 < s_{K\pi} < 2.5\text{ GeV}^2$), illustrating the CP sign reversal driven by cross-channel interference between the $S$-wave $\overline{K}_0^{*0}$ and $P$-wave $\rho^0$ resonances. The red rectangular frames indicate the two representative sub-regions, $\Omega_-$ and $\Omega_+$, characterizing the large negative and positive localized CP asymmetries listed in Table~\ref{tab:integrated_acp}.}
	\label{fig:dalitz_acp}
\end{figure}

Following this isolated scheme, the numerical distribution of the localized direct $CP$ asymmetry driven by the cross-channel interference between $\overline{K}_0^{*0}$ and $\rho^0$ is displayed in Fig.~\ref{fig:dalitz_acp}.

Distinct from a naive rectangular grid, Fig.~\ref{fig:dalitz_acp} explicitly incorporates the curved kinematic boundary of the three-body decay. Specifically, the blank area near the lower-left corner (around $s_{\pi\pi} \sim 0.4 \text{ GeV}^2$ and $s_{K\pi} \sim 1.5 \text{ GeV}^2$) corresponds to the kinematically forbidden region of the Dalitz plot, where the three-body decay is physically impossible. This visual presentation mirrors the mathematical configuration of the integration domain $\Omega$ described above, ensuring that only physically allowed phase space configurations contribute to our phenomenological analysis. The pronounced spatial variations of the localized $CP$ violation across the intersection region, indexed by the color gradient scale on the right, provide critical insights into the underlying hadronic dynamics.

\begin{table}[htbp]
	\centering
	\caption{Localized integrated $CP$ asymmetries $A_{CP}^{\Omega}$ obtained from various two-dimensional phase space regions $\Omega$ in the cross-channel interference intersection between $\overline{K}_0^{*0}$ and $\rho^0$. The theoretical uncertainties are of the same two types as described in Table~\ref{tab:individual_modes}.}
	\label{tab:integrated_acp}
	\renewcommand{\arraystretch}{1.5} 
	\setlength{\tabcolsep}{6mm}      
	\begin{tabular}{ccc}
		\hline\hline
		Range of $s_{\pi\pi}$ ($\text{GeV}^2$) & Range of $s_{K\pi}$ ($\text{GeV}^2$) & $A_{CP}^{\Omega}$ \\ 
		\hline
		[0.39, 0.85] & [1.33, 2.87] & $0.1187_{-0.0316}^{+0.0169}{}_{-0.0112}^{+0.0079}$ \\ \hline
		[0.49, 0.72] & [1.66, 2.43] & $0.2339_{-0.0359}^{+0.0250}{}_{-0.0172}^{+0.0147}$ \\ \hline
		[0.60, 0.80] & [1.50, 2.00] & $-0.3898_{-0.0312}^{+0.0122}{}_{-0.0368}^{+0.0175}$ \\ \hline
		[0.60, 0.80] & [2.00, 2.50] & $0.5109_{-0.0592}^{+0.0279}{}_{-0.0416}^{+0.0142}$ \\ \hline\hline
	\end{tabular}
\end{table}

Owing to the spatial variations of the local $CP$ violation within the overlapping domain, we evaluate the integrated asymmetries $A_{CP}^{\Omega}$ across several representative regions in Table~\ref{tab:integrated_acp}. Within the full-width ($m_R \pm \Gamma$) and half-width ($m_R \pm \frac{1}{2}\Gamma$) resonance intersections, the integrated asymmetries yield $0.1187_{-0.0316}^{+0.0169}{}_{-0.0112}^{+0.0079}$ and $0.2339_{-0.0359}^{+0.0250}{}_{-0.0172}^{+0.0147}$, respectively. To examine the localized sign reversal within the core window, we partition the domain above the $\rho^0$ pole ($s_{\pi\pi} \in [0.60, 0.80]\text{ GeV}^2$) into two symmetric sub-regions: $\Omega_-$ ($s_{K\pi} \in [1.50, 2.00]\text{ GeV}^2$) and $\Omega_+$ ($s_{K\pi} \in [2.00, 2.50]\text{ GeV}^2$), as delineated by the red rectangular frames in Fig.~\ref{fig:dalitz_acp}. The predicted asymmetry shifts remarkably from $A_{CP}^{\Omega_-} = -0.3898_{-0.0312}^{+0.0122}{}_{-0.0368}^{+0.0175}$ to $A_{CP}^{\Omega_+} = +0.5109_{-0.0592}^{+0.0279}{}_{-0.0416}^{+0.0142}$. This sign reversal is qualitatively consistent with the rapid transition from the negative bins (blue region, $\sim -0.5$) to the positive bins (red region, $\sim +0.6$) observed in the LHCb data (Fig.~\ref{fig:dalitz_zoom}), supporting the cross-channel interference dynamics.

The pronounced spatial fluctuations of the localized CP asymmetry $A_{\mathrm{CP}}^\Omega$, along with its distinct sign reversal across the intersection region, manifest a cooperative mechanism between the rapid evolution of the relative strong phase $\Delta\delta(s_{\pi\pi},s_{K\pi})$ and the dramatic suppression of the total decay width in the denominator. Crucially, inside the localized subregions where the CP asymmetry approaches unity, the underlying physics does not stem from an absolute surge in the production of one charge channel (e.g., $B^+$); rather, it is driven by a specific strong phase configuration that induces a near-complete destructive interference in the conjugate charge channel (e.g., $B^-$), suppressing its local differential width toward an extreme minimum near zero. Under this circumstance, the total differential decay rate (the denominator) is severely depressed by the depletion of one conjugate channel, allowing even a small absolute difference (or a suppressed local differential branching fraction) in the numerator to be relatively amplified into a pronounced localized CP asymmetry close to $\pm 1$. Consequently, the CP asymmetry exhibits a sweeping trajectory from $-0.7126$ to $+0.7308$ within an extremely narrow kinematic region.Generally, such delicate microscopic interference fringes, which heavily rely on denominator minimization, are highly susceptible to being washed out and masked in the total phase space by overwhelming non-resonant backgrounds or dominant resonances with much larger branching fractions. Fortunately, because both the $\overline{K}_0^{*0}$ and $\rho^0$ components possess sufficiently large individual branching fractions, their cross-channel interference pattern successfully overcomes such masking and stands out with high resolution as a prominent local feature on the global Dalitz plot. As illustrated in Fig.~\ref{fig:dalitz_acp}, the cross-channel interference between the $\overline{K}_0^{*0}$ and $\rho^0$ amplitudes yields a substantial positive contribution to the local CP asymmetry. This dominant mechanism successfully explains the absence of the expected typical $\rho^0$ pole-crossing CP sign reversal within the localized phase-space region defined by $0.4 \text{ GeV}^2 < s_{\pi\pi} < 0.8 \text{ GeV}^2$ and $1.5 \text{ GeV}^2 < s_{K\pi} < 2.5\,\mathrm{GeV}^2$.

\section{Conclusion}
In summary, we have elucidated the dynamical origin of the localized direct $CP$ violation patterns in the charmless three-body decay $B^{-}\rightarrow K^{-}\pi^{+}\pi^{-}$ by focusing on the cross-channel $S$-$P$ wave interference. By implementing a quasi-two-body factorization scheme that couples short-distance PQCD hard kernels with long-distance $S$-wave LASS and $P$-wave RBW form factors, we have mapped the fine quantum coherent structure within the overlapping resonance region. Our findings demonstrate that the long-standing puzzle of the pronounced rightward tilt in the $CP$ sign-reversal pattern is not an isolated fluctuation, but a natural consequence of the orthogonal geometric topology between the scalar $\overline{K}_0^{*0}$ and vector $\rho^{0}$ resonance bands on the Dalitz plot.

Crucially, this work provides a clearer physical perspective on the emergence of large local $CP$ asymmetries. Rather than originating from an absolute enhancement of the decay amplitude in a single charge channel, these pronounced fluctuations are driven by a specific strong phase configuration that triggers substantial destructive interference in the conjugate charge channel. This denominator-minimization mechanism offers a clear and generalizable phenomenological framework, illustrating how modest variations in the interference numerator can be significantly amplified within narrow kinematic boundaries.

The fact that these microscopic interference fringes successfully emerge on the global Dalitz plot underlines the prominent role of cross-channel dynamics in channels with sufficiently large intermediate branching fractions. Consequently, our results underscore a notable methodological limitation in traditional experimental analyses: standard isobar models, which rely on empirical additive frameworks with freely varied fitting parameters, cannot fully account for the underlying dynamical origin of the rapid phase transitions in overlapping resonance regions. Instead of introducing unconstrained empirical parameters, incorporating proper cross-channel $S$-$P$ wave coherent interactions derived from short-distance QCD factorization constraints is important for improving the description and fitting quality of local hadronic distributions. Looking forward to the upcoming high-luminosity era at Belle II~\cite{Belle-II:2018jsg} and the High-Luminosity LHC~\cite{LHCb:2018roe}, the impending abundance of high-statistics data will elevate these subtle interference patterns into high-precision benchmarks for new physics searches and the clean extraction of the CKM weak phase $\gamma$.

 \section*{Acknowledgements}
This work was supported by National Natural Science
Foundation of China under Grants No. 12475096.


\end{spacing}

\begin{thebibliography}{99}
	
	\bibitem{Cheng:2007shb}
	H.-Y. Cheng, C.-K. Chua, and A. Soni, Phys. Rev. D \textbf{76}, 094006 (2007), arXiv:0704.1049 [hep-ph].
	
	\bibitem{Klein:2017xti}
	R. Klein, T. Mannel, J. Virto, and A. Khodjamirian, Phys. Rev. D \textbf{95}, 093009 (2017), arXiv:1703.00259 [hep-ph].
	
	\bibitem{Belle:2005rpz}
	A. Garmash \textit{et al.} (Belle Collaboration), Phys. Rev. Lett. \textbf{96}, 251803 (2006), arXiv:hep-ex/0512066.
	
	\bibitem{LHCb:2019sus}
	R. Aaij \textit{et al.} (LHCb Collaboration), Phys. Rev. D \textbf{101}, 012006 (2020), arXiv:1909.05212 [hep-ex].
	
	\bibitem{BaBar:2008lpx}
	B. Aubert \textit{et al.} (BaBar Collaboration), Phys. Rev. D \textbf{78}, 012004 (2008), arXiv:0803.4451 [hep-ex].
	
	\bibitem{Belle:2004drb}
	A. Garmash \textit{et al.} (Belle Collaboration), Phys. Rev. D \textbf{71}, 092003 (2005), arXiv:hep-ex/0412066.
	
	\bibitem{LHCb:2013fio}
	R. Aaij \textit{et al.} (LHCb Collaboration), Phys. Rev. Lett. \textbf{111}, 101801 (2013), arXiv:1306.1246 [hep-ex].
	
	\bibitem{Cheng:2020iwx}
	H.-Y. Cheng and C.-K. Chua, Phys. Rev. D \textbf{102}, 053006 (2020), arXiv:2007.02534 [hep-ph].
	
	\bibitem{LHCb:2022fpg}
	R. Aaij \textit{et al.} (LHCb Collaboration), Phys. Rev. D \textbf{108}, 012008 (2023), arXiv:2206.07622 [hep-ex].
	
	\bibitem{LHCb:2019jta}
	R. Aaij \textit{et al.} (LHCb Collaboration), Phys. Rev. Lett. \textbf{124}, 031801 (2020), arXiv:1909.05211 [hep-ex].
	
	\bibitem{Bhattacharya:2013boa}
	B. Bhattacharya, M. Imbeault, and D. London, Phys. Rev. D \textbf{89}, 054013 (2014), arXiv:1303.0046 [hep-ph].
	
	\bibitem{Cabibbo:1963yz}
	N. Cabibbo, Phys. Rev. Lett. \textbf{10}, 531 (1963).
	
	\bibitem{Giri:2003ty}
	A. Giri, Y. Grossman, A. Soffer, and J. Zupan, Phys. Rev. D \textbf{68}, 054018 (2003), arXiv:hep-ph/0303187.
	
	\bibitem{Dedonder:2010fg}
	J. P. Dedonder, L. Lesniak, and R. Thomas, Phys. Rev. D \textbf{84}, 074033 (2011), arXiv:1011.0960 [hep-ph].
	
	\bibitem{HFLAV:2022pvk}
	Y. S. Amhis \textit{et al.} (HFLAV Collaboration), Phys. Rev. D \textbf{107}, 052008 (2023), arXiv:2206.07501 [hep-ex].
	
	\bibitem{Wang:2020gqm}
	W.-F. Wang, H.-n. Li, and C.-D. L\"u, Phys. Rev. D \textbf{103}, 036003 (2021), arXiv:2011.04212 [hep-ph].
	
	\bibitem{Zhou:2023dcl}
	S.-H. Zhou, X.-X. Hai, R.-H. Li, and C.-D. Lu, Phys. Rev. D \textbf{107}, 116023 (2023), arXiv:2303.09062 [hep-ph].
	
	\bibitem{Chai:2022prb}
	J. Chai, S. Cheng, and A.-J. Ma, Phys. Rev. D \textbf{105}, 033003 (2022), arXiv:2111.05047 [hep-ph].
	
	\bibitem{Hua:2020usv}
	J. Hua, H.-n. Li, C.-D. Lu, W. Wang, and Z.-P. Xing, Phys. Rev. D \textbf{104}, 016025 (2021), arXiv:2012.15074 [hep-ph].
	
	\bibitem{Chen:2002th}
	C.-H. Chen and H.-n. Li, Phys. Lett. B \textbf{561}, 258 (2003), arXiv:hep-ph/0209043.
	
	\bibitem{Zou:2020fax}
	Z.-T. Zou, Y. Li, and X. Liu, Eur. Phys. J. C \textbf{80}, 517 (2020), arXiv:2005.02097 [hep-ph].
	
	\bibitem{Cheng:2013dua}
	H.-Y. Cheng and C.-K. Chua, Phys. Rev. D \textbf{88}, 114014 (2013), arXiv:1308.5139 [hep-ph].
	
	\bibitem{Belle:2019rup}
	Y. T. Lai \textit{et al.} (Belle Collaboration), Phys. Rev. D \textbf{100}, 011101 (2019), arXiv:1904.06835 [hep-ex].
	
	\bibitem{Aston:1987ir}
	D. Aston \textit{et al.}, Nucl. Phys. B \textbf{296}, 493 (1988).
	
	\bibitem{BaBar:2005qms}
	B. Aubert \textit{et al.} (BaBar Collaboration), Phys. Rev. D \textbf{72}, 072003 (2005), [Erratum: Phys. Rev. D \textbf{74}, 099903 (2006)], arXiv:hep-ex/0507004.
	
	\bibitem{Wang:2020saq}
	W.-F. Wang, J. Chai, and A.-J. Ma, JHEP \textbf{03}, 162 (2020), arXiv:2001.00355 [hep-ph].
	
	\bibitem{Zhang:2013oqa}
	Z.-H. Zhang, X.-H. Guo, and Y.-D. Yang, Phys. Rev. D \textbf{87}, 076007 (2013), arXiv:1303.3676 [hep-ph].
	
	\bibitem{Zhang:2013iga}
	Z.-H. Zhang, X.-H. Guo, and Y.-D. Yang, arXiv:1308.5242 [hep-ph].
	
	\bibitem{Wang:2015ula}
	C. Wang, Z.-H. Zhang, Z.-Y. Wang, and X.-H. Guo, Eur. Phys. J. C \textbf{75}, 536 (2015), arXiv:1506.00324 [hep-ph].
	
	\bibitem{Bediaga:2006jk}
	I. Bediaga, G. Guerrer, and J. M. de Miranda, Phys. Rev. D \textbf{76}, 073011 (2007), arXiv:hep-ph/0608268.
	
	\bibitem{Keum:2000wi}
	Y.-Y. Keum, H.-n. Li, and I. Sanda, Phys. Rev. D \textbf{63}, 054008 (2001), arXiv:hep-ph/0004173.
	
	\bibitem{Lu:2000em}
	C.-D. L\"u, K. Ukai, and M.-Z. Yang, Phys. Rev. D \textbf{63}, 074009 (2001), arXiv:hep-ph/0011142.
	
	\bibitem{Cheng:2014pqa}
	S. Cheng and Z.-J. Xiao, Front. Phys. \textbf{9}, 568 (2014).
	
	\bibitem{Xiao:2022ebt}
	Z.-J. Xiao and X. Liu, Chin. Phys. C \textbf{46}, 123103 (2022), arXiv:2207.04190 [hep-ph].
	
	\bibitem{Li:1994iu}
	H.-n. Li, Phys. Rev. D \textbf{52}, 3958 (1995), arXiv:hep-ph/9405259.
	
	\bibitem{Keum:2000ph}
	Y.-Y. Keum and H.-N. Li, Phys. Rev. Lett. \textbf{86}, 4215 (2001), arXiv:hep-ph/0004059.
	
	\bibitem{Yang:2022ebu}
	Y. Yang, X. Zhao, L. Lang, J. Huang, and J. Sun, Chin. Phys. C \textbf{46}, 083103 (2022), arXiv:2201.12834 [hep-ph].
	
	\bibitem{Ali:1998eb}
	A. Ali, G. Kramer, and C.-D. Lu, Phys. Rev. D \textbf{58}, 094009 (1998), arXiv:hep-ph/9804363.
	
	\bibitem{Buchalla:1995vs}
	G. Buchalla, A. J. Buras, and M. E. Lautenbacher, Rev. Mod. Phys. \textbf{68}, 1125 (1996), arXiv:hep-ph/9512380.
	
	\bibitem{Shen:2006ms}
	Y.-L. Shen, W. Wang, J. Zhu, and C.-D. Lu, Eur. Phys. J. C \textbf{50}, 877 (2007), arXiv:hep-ph/0610380.
	
	\bibitem{ParticleDataGroup:2024cfk}
	S. Navas \textit{et al.} (Particle Data Group), Phys. Rev. D \textbf{110}, 030001 (2024).
	
	\bibitem{Belle-II:2018jsg}
	W. Altmannshofer \textit{et al.} (Belle-II Collaboration), PTEP \textbf{2019}, 123C01 (2019), [Erratum: PTEP \textbf{2020}, 029201 (2020)], arXiv:1808.10567 [hep-ex].
	
	\bibitem{LHCb:2018roe}
	R. Aaij \textit{et al.} (LHCb Collaboration), arXiv:1808.08865 [hep-ex].
	
\end{thebibliography}
\end{document}